\documentclass[twocolumn]{aastex631}

\usepackage{xcolor}

\usepackage{bm}
\usepackage{amsmath}
\usepackage{empheq}
\usepackage{enumitem}
\usepackage{soul}

\newcommand{\e}{\operatorname{e}}
\newcommand{\I}{\operatorname{i}}

\newcommand{\rbar}{\operatorname{cov}}

\def\rbar{\bar \rho}
\def\nbar{\bar n}
\def\msol{{\rm M}_\odot}

\def\vrms{V_{\rm rms}}
\def\kms{{\rm km}\, {\rm s}^{-1}}
\def\cc{{\rm cm}^{-3}}

\shorttitle{Initial mass function and star formation rate in the CMZ}
\shortauthors{Chabrier \& Dumond }
\graphicspath{{./}{figures/}}

\begin{document} 

   \title{
   A consistent explanation for the unusual  initial mass function and star formation rate in the Central Molecular Zone (CMZ) 
   }
%   \titlerunning{Initial mass function and star formation rate in the Central Molecular Zone}

   % \subtitle{}

\author{Gilles Chabrier}
\affiliation{Ecole normale sup\'erieure de Lyon, CRAL, Universit\'e de  Lyon, UMR CNRS 5574, F-69364 Lyon Cedex 07, France}
\affiliation{School of Physics, University of Exeter, Exeter, EX4 4QL, UK}
   
\author{Pierre Dumond}
\affiliation{Ecole normale sup\'erieure de Lyon, CRAL, Universit\'e de  Lyon, UMR CNRS 5574, F-69364 Lyon Cedex 07, France}

\email{chabrier@ens-lyon.fr ;   pierre.dumond@ens-lyon.fr}

   \date{Received XXX; accepted XXX}

  \begin{abstract}
We examine various physical processes that may explain the shallow high-mass slope of the IMF as well as the low SFR in star-forming molecular clouds (MCs) in the Central Molecular Zone (CMZ). 
We show that the strong tidal field and the tidal shear experienced by the CMZ have opposite effects on the collapse of density fluctuations and nearly compensate, but in any case have a negligible impact and can not explain these unusual properties.
Similarly, we show that the intense magnetic field in the CMZ provides a negligible pressure support and, for the high densities at play should not modify the probability density function (PDF) of the turbulent gas flow in the clouds, thus affecting negligibly the slope of the IMF. 
However, we show that, in contrast to MCs in the Galactic disk, the ones in the CMZ experience only one single episode of turbulence injection at large scale, most likely due dominantly to bar gas inflow. Indeed, their rather short lifetime, due to their high mean densities, is similar to one typical turbulence crossing time. Consequently, according to the Hennebelle-Chabrier theory of star formation, within this 'single turbulence episode' scenario, the cloud experiences one single field of turbulence induced density fluctuations, leading eventually to gravitationally unstable prestellar cores. As shown in \cite{HennebelleChabrier2013}, this yields a flatter IMF than usual and leads to the correct observed slope for the CMZ star-forming clouds. Similarly, this single large scale turbulence event within the cloud lifetime yields a 5 to 6 lower SFR than under usual MW cloud conditions, again in agreement with the observed values. Therefore, we suggest that this 'single large scale turbulence injection' episode can explain both the shallow IMF high-mass slope and low SFR of clouds in the CMZ. 
  \end{abstract}

%
%-------------------------------------------------------------------

\section{Introduction}
\label{intro}

The Central Molecular Zone (CMZ), i.e. the region within a Galactocentric radius $R\simeq 300$ pc, hosts several young massive clusters, among which the young nuclear cluster (YNC; $\sim$2.5-5.8 Myr, $M\ge 2\times 10^4\,\msol$) \citep{Lu+2013}  and the Arches cluster ($\sim$2-4 Myr, $M\sim 4-6\times 10^4\,\msol$)  \citep{Lohr+2018}
%Walker+2021: the Brick (IRDC): $\bar{n} > 10^4\cc, M>10^4\msol, L\sim 2-3$ pc; (e.g. Immer et al. 2012; Longmore et al. 2012; Walker et al. 2015); $T_{dust}\sim T_{gas}>\sim 120-160$ K (\S3.6.1); Discussion: $M_J=0.35$-2.15 $\msol$ 
%$V_{rms}\approx 10-20\kms$
characterised by an unusual stellar initial mass function (IMF) compared with the canonical IMF, which is usually found to exhibit very little variations in various environments. Indeed, the slopes of these IMFs, $\mathcal{N}(M)=dN/dM\propto M^{-\alpha}$,
have been found to be significantly flatter than the usual Salpeter slope ($\alpha = 2.35$), with $\alpha = 1.7\pm 0.2$ for the YNC \citep{Lu+2013} and 
$\alpha = 1.80^{0.05}_{0.05}$-$2.0^{0.14}_{0.19}$ for the Arches, depending on the functional form fitted, 
with potentially a strong steepening above $\sim4$-8 $\msol$ \citep{Hosek+2019}. The same behaviour, with $\alpha = 1.68^{0.13}_{0.09}$ has been found for the Quintuplet cluster \citep{Hussman2012}, although it is unclear whether or not this is due to mass segregation for this slightly older cluster. Other Galactic clusters might have slopes consistent with that seen in the Arches, e.g. Wd1, NGC3603  \citep{Pang+2013, Lim+2013, Andersen+2017,Lu+2020}, although these results are more uncertain. There is also tentative evidence that this shallow mass function is also present for the prestellar cores, with an apparent excess of high-mass sources, even though substantial uncertainties remain in these measurements \citep[]{Lu+2020,Henshaw+2023}. 

Furthermore, the mean value of the star formation rate (SFR) in the CMZ is found to be about $ 0.07^{0.08}_{0.02} \,\msol\,{\rm yr}^{-1}$, significantly below the Kennicutt-Schmidt relation for its gas surface density and about an order of magnitude below Lada's relationship \citep{Lada+2012} between the SFR and the mass of dense, molecular gas, a relationship which holds for nearby molecular clouds as well as for external galaxies \citep{Longmore+2013,Hosek+2019}. It is not clear yet, however,
whether this low SFR for the amount of dense gas is also found for the CMZs of external galaxies or only for the Galactic CMZ.
Numerous studies have been devoted to the unusual SFR of the CMZ, invoking, in particular, feedback radiation or Galactic shear to explain its lower than expected value \cite[see][]{Henshaw+2023}, with no real conclusive explanation. To the best of our knowledge, however, only one study has been devoted
to the puzzle of the top heavy, shallow IMF of the CMZ clusters, though only through numerical simulations \citep{Dib+2007}. 

In this paper, we examine both issues, IMF and SFR, throughout an analytical exploration of the various physical processes that could be responsible for these CMZ particular properties. The paper is organised as follows. In \S\ref{sec-CMZ}, we summarise the thermodynamic and dynamical properties of the CMZ. In \S\ref{sec-imf}, we recall the various relevant scaling properties for star-forming molecular clouds (MCs). 
In \S\ref{sec-tidal} and \ref{sec-shear}, we derive the equations allowing to characterise the impact of the Galactic tidal field and shear upon the IMF, respectively. In  \S\ref{sec-dynamics}, we
examine the dynamics of the star formation process in the CMZ and explore the consequences on both the IMF and the SFR. %\S\ref{sec-sfr}. 
Section \ref{conclusion} is devoted to the conclusion.

\section{Properties of the Central Molecular Zone}
\label{sec-CMZ}

A recent summary of the properties of the CMZ can be found in the review of Henshaw et al. (2023). The conditions in the CMZ are extreme compared with the Solar neighbourhood and comparable to those found in the early universe \cite[e.g.,][]{KruijssenLongmore2013}. The part where most of the molecular gas is located, $45\lesssim r/{\rm pc}\lesssim 115$, denoted the gas stream, includes all the major cloud complexes in the Galactic centre (GC).
%The flow of matter in the CMZ is controlled by the gravitational potential of the bar

The molecular gas in the CMZ is distributed along an approximately ring like, or possibly a more torus-like structure \citep{Kruijssen+2015}, and thus follows eccentric orbits. This gas exhibits velocity dispersions much larger than those measured in Galactic disk clouds, indicating a much higher level of turbulence, as found in extreme environments. Various mechanisms can be responsible for injecting such amounts of large-scale turbulence, a point we will examine in more details in \S\ref{sec-turb}.
%supernova feedback or bar instabilities seem to be the dominant contributors to this process, depending on the location \cite[see e.g.][]{Krumholz+2017}.

The properties of the CMZ are the following \citep{Henshaw+2023}. The temperature ranges from about 50 to 100 K which yields a typical sound speed $C_S\simeq 0.4$-0.6 $\kms$. The scale dependence of the mean density of MCs in the CMZ can be inferred from the determinations of \cite{Tanaka+2020} and \cite{Krieger+2020}:
\begin{equation}
    \label{scaling_density}
    M=M_0\left(\frac{R}{1\,\text{pc}}\right)^{2.7}\,\msol \implies \bar{n} =n_0\left(\frac{R}{1\,\text{pc}}\right)^{-0.3}\,\cc.
\end{equation}
However, while the first authors suggest $M_0=10^{3.5}\,\msol$,  yielding $n_0=6\times 10^4\,\cc$, the second group gets $M_0=200\,\msol$, then $n_0=3.8\times 10^3\,\cc$. Typical sizes for star-forming MCs in the CMZ are $L_c\sim1$-10 pc.
%\(10^{4}\) to \(10^{6}\) cm\(^{-3}\). 
The scaling properties of the 3D rms velocity slightly differ from the usual Larson relation, with 
\begin{equation}
    \label{scaling_velocity}
\vrms(R) \equiv \vrms^{3D}(R)={\sqrt 3}\vrms^{1D}(R)={\sqrt 3} V_0\times \left(\frac{R}{1\,{\rm pc}}\right)^{\eta}, %\,\,\,\,\,{\rm with }\,\,\eta=0.7 
\end{equation}
where $V_0$ is the 1D observational determination along the line of sight. Whereas the power index $\eta\simeq 0.7$ is found in most studies, the observationally determined normalization at 1 pc differs substantially. It should be stressed, however, that these determinations do not probe the same scales! Using tracers CO(3-2) and CO(1-0) from 1 pc to 100 pc,
\cite{Krieger+2020} find a  value $V_0=2\,\kms$. At smaller scales, in the range 0.1 to 1 pc, \cite{Tanaka+2020}, using HCN as a tracer, find $V_0=10\,\kms$.  
At even smaller scales, \cite{Henshaw+2019}, using a different method to determine the velocity dispersion, find an even larger
value for the Brick, $V_0(0.07\,{\rm pc})=4.4\pm 2.2\,\kms$, substantially higher than extrapolations from the above larger scale determinations. Thus, there seems to be an increase of the observed velocity  dispersion normalisation with decreasing scale. It is worth noting that while 1 to 100 pc scales are relevant for {\it cloud} scales, the smaller ($\lesssim 1$ pc) scales are more representative of {\it core} scales. As will be discussed later, it is important to take into consideration this difference of scales in the observational determination of the velocity dispersion, depending on whether one considers clouds or prestellar cores, which is the case for the IMF/CMF.
%Therefore, the normalization of \cite{Tanaka+2020} makes more sense when studying the core mass function (CMF).
We will come back to the issue in \S\ref{shear_dissipation}. In the study of \citet{Federrath+2016}, devoted to one of the CMZ clouds, the turbulence appears to be purely solenoidal, implying a factor $b\simeq 1/3$ or even  $b=0.2$ for the turbulence driving parameter, a consequence of the strong shear. This does not imply, however, that this dominant solenoidal nature of turbulence applies to all clouds in the CMZ.

Using eqns.(\ref{scaling_density}) and (\ref{scaling_velocity}), we can determine a few typical mean quantities for the CMZ (taking a fiducial value $\nbar=10^4\,\cc$), notably the Jeans length, $ \lambda_J=\sqrt{C_s/G\rbar}\simeq 0.4$-0.6 pc, and Jeans mass \(M_J=(4\pi/3)\lambda_J^3\rbar\approx 50\)-150 $\msol$.  The
typical 3D rms Mach number at the injection scale ($L_i\simeq L_c$ for the star-forming clouds) lies in
the range $\mathcal{M}={\sqrt 3}\frac{V_0}{C_s}\left(\frac{L_i}{1\text{pc}}\right)^{\eta}\approx 8$-170, depending on the chosen 1-pc normalization value. This yields cloud typical turbulent crossing time and sonic length:
\begin{eqnarray}
\tau_{ct}&\simeq& {L_c\over \vrms^{1D}(L_c)} \approx 0.2-2 \,{\rm Myr}\,\,\,\,\,{\rm for }\,\,\,\,L_c\sim1-10 \,{\rm pc}\\
l_s&=&({C_s\over V_0})^{1/\eta}=\mathcal{M}^{-1/\eta}\,L_c\approx 0.01-0.1\,{\rm pc}. %\,\,\,\,{\rm or}\,\,\,\,\approx 2-20\,{\rm pc}.
    \label{tct}
\end{eqnarray}
 
\section{The slope of the IMF}
\label{sec-imf}

\subsection{Impact of the velocity-size relation}
\label{sec-velocity-size}

In the general gravoturbulent scenario of star formation \citep{PadoanNordlund2002,HennebelleChabrier2008,Hopkins2012}, density fluctuations in MCs are due to the shock cascade of large scale compressible turbulence characterised by a log-density power spectrum ${\mathcal P}(\delta) \propto k^{-n}$, where $\delta=\log(\rho/\rbar)$, with a typical 3D index found to be very similar to the one found for the velocity, $n\simeq 3.8$ \citep{Beresnyak+2005}. The variance of the dispersion of the log-density field generated by turbulence at scale $R$ is taken to be \citep{HennebelleChabrier2008} (HC08)
\begin{equation}
   \sigma^2(R)=\sigma_0^2\,[1-({R\over L_i})^{n-3}],
   \label{eqn-imf}
\end{equation} 
The variance at small (core) scale ($R\ll L_c$), which is what really matters for the IMF, is found in hydrodynamical
and MHD simulations to be reasonably well described {\it under some conditions} by the relation \citep{Molina+2012}:
\begin{equation}
\sigma_0^2=\ln\left[1+(b\mathcal{M})^2(\frac{\beta}{1+\beta})\right],
\label{variance}
\end{equation}
where $\beta=2(\mathcal{M}_A/\mathcal{M})^2$, $\mathcal{M}_A=V_{rms}/V_A$ is the 3D rms Alfvenic Mach number, where $V_A=B_{3D}/{\sqrt{\mu_0\rbar}}$ denotes the  3D mean Alfv{\'e}n velocity with $\mu_0$ the permeability of vacuum.

In the canonical Hennebelle \& Chabrier theory of the IMF (eqn.(29) of HC08), the collapsing barrier for the density fluctuations of scale $R$ generated by turbulence reads:
\begin{equation}
   e^{s^c_R} = \frac{\rho^c_R}{\rbar}\ge  \tilde{R}(1+\mathcal{M}_\star^2 \tilde{R}^{2\eta}), 
 \label{eqn-HC}
\end{equation}
which yields for the dominant term in the IMF 
\begin{equation}
   \mathcal{N}({M_R^c}) = \frac{dN}{d{\tilde M}_R^c} \propto \frac{1}{{\tilde M}_R^c}  \frac{d\e^{s_R^c}}{d{\tilde R}} \frac{d{\tilde R}}{d{\tilde M}_R^c}{\cal P}(\e^{s_R^c}), 
 \label{eqn-imfHC}
\end{equation}
where $\tilde{R}=R/\lambda_J$ and $\tilde{M}=M/M_J$ are  normalized to the Jeans length  and the Jeans mass, respectively. $\mathcal{M}_\star=\frac{\mathcal{M}}{\sqrt{3}}\left(\frac{\lambda_J}{L_i}\right)^\eta \simeq ({\lambda_J\over 1\,{\rm pc}})({V_0\over C_S})\approx 2$ and $\sim$10 for \cite{Krieger+2020} and \cite{Tanaka+2020} 1 pc normalizations, respectively, is the Mach number at the Jeans scale. In the time-dependent extension of the theory
(eqn.(21) of \cite{HennebelleChabrier2013}, HC13) the slope of the high-mass tail of the IMF is given by:
\begin{equation}
    \alpha=\frac{4+2\eta}{2\eta+1}+6\frac{\eta-1}{(2\eta+1)^2}\frac{\ln(\mathcal{M}_\star)}{\sigma^2}.
\label{eqn-imf}
\end{equation}
One can already notice that, because of the larger than usual Larson turbulent index $\eta=0.7$ (instead of $\eta\approx 0.5$), the first term in eqn.(\ref{eqn-imf}) yields a  shallower slope than the one predicted in the solar neighbourhood by a factor $\sim 0.25$ while, under the present conditions, the second term amounts to about -0.1.
%As emphasized in \cite{Chabrier_VariationsStellarInitial2014},  is negligible in highly turbulent and dense regimes because of the large variance \(\sigma^2\).
This flattening reflects the high gas dispersion   due to more vigorous turbulence which prevents large-scale star forming clumps from collapsing \citep{ChabrierHennebelle2011}. 
%The first term  in eqn.\ref{eqn-imf} is equal to 2.25 instead 2.5 with \(\eta=0.5\). Moreover,  . 
Then, although contributing to the flattening of the high-mass tail of the IMF, this modification of the  velocity-size relation of turbulence is not sufficient to explain the aforementioned observed values of $\alpha$.

\subsection{Impact of gravity on the PDF}
\label{sec-pdf}
It must be remembered that the HC theory is based on the assumption of a purely lognormal probability density function (PDF) in MCs. There is ample observational evidence, however, that
the PDF of star-forming clouds develops a power law at high density, due to the onset of gravity, and thus cannot be entirely described by a lognormal form
\citep[e.g., ][]{Schneider+2022}.  Recently, \cite{JaupartChabrier2020} developed a {\it predictive} analytical theory to characterize the impact of gravity on the PDF of turbulence. They demonstrated that the power law will start to develop above a critical density $s_{\rm crit}=\log({\rho_{\rm crit}} /\rbar)$ given by the condition:
\begin{equation}
    e^{s_{\rm crit}} = \frac{45b^2\mathcal{M}_\star^2}{\pi\sigma_0^2}(s+\frac{1}{2}\sigma_0^2), 
\end{equation}
assuming that \(\sigma_0\) weakly depends on  scale. This yields $s_{\rm crit}\simeq 9$ for the value of $V_0$ obtained by  \cite{Tanaka+2020} and $s_{\rm crit}\simeq 5$ for the value derived by \cite{Krieger+2020}, respectively. The fact that no (or a quite small) power law is observed in the PDF of MCs of the CMZ \citep{Henshaw+2023} suggests that $V_0$ should be close to the first value. This is also consistent with the higher velocity amplitude suggested by \cite{Henshaw+2019} for the Brick. The Jaupart-Chabrier formalism thus shows that the power law in the PDF of the CMZ will develop at a higher density (or later for a given high enough density) than for standard Milky Way MCs, highlighting again the vigorous level of turbulence in these regions. 

\subsection{Impact of the magnetic field}
\label{sec-magnetic}

The magnetic field in the CMZ ranges from about 10 to 1000 \(\mu G\), a much stronger value than in the solar neighbourhood \citep{Henshaw+2023,Lu+2023}. It is generally
admitted that at high density the amplitude of the magnetic field scales as \(B\propto \rho^{n_B}\), with $n_B$ varying from 1/2 to 0 at very high density, where it saturates.  We examine both cases below.

In the case $n_B=1/2$, the magnetic energy in the Virial equation does not depend on the scale, so we simply need to rescale the Jeans length as 
\begin{equation}
    C_s^\prime \rightarrow (C_s^2+V_A^2)^{1/2} \ ; \lambda_J^B=\frac{C_s^\prime}{\sqrt{G\bar{\rho}}} \ ; \ \mathcal{M}_\star=\frac{1}{\sqrt{3}}\frac{V_0}{C_s^\prime}\left(\frac{\lambda_J^B}{1\text{pc}}\right)^{2\eta},
\end{equation}
where $V_A=B/{\sqrt{\mu_0\rbar}}\sim 1\,\kms$.

In the case $n_{B}=0$, the impact of the magnetic field pressure can be taken into account by rescaling  $\mathcal{M}_\star$ (see \S2.3.1 of HC13):
\begin{eqnarray}
    \mathcal{M}_\star^2\rightarrow\frac{\mathcal{M}_\star^2}{2}\left(1+\sqrt{1+\frac{4{V_A^\star}^2}{\mathcal{M}_\star^4}}\right) \\
     \text{with} \ {V_A^\star}=\frac{1}{{\sqrt 6}}\frac{V_A}{C_s}=\frac{1}{{\sqrt 6}}\frac{\mathcal{M}}{\mathcal{M}_A}.
\end{eqnarray}
The collapsing barrier is the same as eqn.(\ref{eqn-HC}) with the rescaled value of $\mathcal{M}_\star$.

Typical CMZ conditions yield ${\mathcal{M}_A}\sim 30,  V_A^\star\sim1$. For the velocity dispersions of \cite{Tanaka+2020}, the correction on $\mathcal{M}_\star$, thus on the slope, is found to be negligible. For the smaller value of \cite{Krieger+2020}, the correction on the slope is about $\sim 0.05$.
% Recent measurements of the magnetic field in CMZ clouds \citep{Lu+2023} suggest sightly smaller magnetic field values, about 0.5 \(\mu G\), i.e. ${\mathcal{M}_A}\sim 1,  V_A^\star\sim 5$. 
In all cases, the impact of the magnetic field pressure on the slope of the IMF is found to be too small to explain the flattening of the high-mass tail of the IMF.

It is interesting to estimate the magnetic field that would be necessary to explain the flattening of the high-mass slope of the IMF, $\Delta \alpha \sim 0.6$ (see \S\ref{intro}). Using eqn.(\ref{eqn-imf}), the slope difference between the magnetized and non-magnetized ($B=0$) cases is:
\begin{equation}
    \alpha_B-\alpha_{B=0}={6(\eta -1)\over (2\eta+1)^2 \sigma^2} \ln ({\mathcal{M}_\star \over {\mathcal{M}_\star^{B=0}}}) \approx -\frac{0.3}{\sigma^2}  \ln ({\mathcal{M}_\star \over \mathcal{M}_\star^{B=0}})
\end{equation}
Picking the most favorable case, i.e. a small (1 pc), moderately turbulent (\citep{Krieger+2020} rms velocity normalization) cloud, for the magnetic pressure to provide sufficient support to explain the slope of the IMF, the magnetic field should be  about 3 orders of magnitude larger than the {\it upper limit} of the aforementioned observed values. Using the velocity normalization of \cite{Tanaka+2020}, the field would need to be at least 5 orders of magnitude larger than this upper value.

However, depending on its dependence upon density,  the magnetic field  can modify or not the PDF of the turbulence in the star-forming gas.
In the case $n_{B}=1/2$,  the magnetic field yields a narrower variance than in the purely hydrodynamical case, as seen from eqn.(\ref{variance}). 
Recent observations \citep{Lu+2023} have allowed to infer the {\it turbulent} component of the magnetic field, $B_t$, in some clouds of the CMZ, 
$V_{A,rms}\equiv V_{A,t}=(\langle B_t^2\rangle / \langle B_{tot}^2\rangle)^{1/2} V_{A,tot}$. Using Table 1 of these authors, we get typical values of the rms Alfvenic 
velocity $V_{A,rms}\approx 1-$3 $\kms$ and  rms Alfvenic Mach numbers $\mathcal{M}_A\approx 20$-40. This yields $\beta\approx 0.1$-0.5. These values are consistent with the results of the numerical simulations of \cite{Federrath+2016}.
In that case, we have verified that the modification of the PDF of the gas due to the strong turbulent magnetic field flattens significantly the slope of the IMF and yields potentially values consistent with the observed value. This will be illustrated later on in \S\ref{sec-dyn-imf}. As will be seen, however, this depends on the chosen value of the field at core scales, a quantity very difficult to determine precisely.

\noindent However, a case $n_{B}=1/2$ in the magnetic field-density dependence is probably not realistic for the high densities typical of the CMZ clouds ($\nbar \gtrsim 10^4\,\cc$). At such densities,  the gas and the magnetic field are decoupled and the shock continuity equation is independent of the magnetic field strength, yielding $n_{B}\simeq 0$ \citep{Molina+2012}. 
In that case, we recover the non-magnetic, hydrodynamical regime ($\beta \rightarrow \infty$) and the density variance in eqn.(\ref{variance}) simply becomes $\sigma_0^2=\ln\left[1+(b\mathcal{M})^2\right]$ \citep{Padoan+1997}.

%Figure \ref{fig-mag} shows the impact of the modification of the slope of the IMF due to the magnetic field for these values.

\section{The Tidal Field}
\label{sec-tidal}

Due to its central Galactic location, the CMZ experiences the strong impact of the Galactic tidal field.  
The spherically symmetric, enclosed mass distribution over the range $R = 1$–300 pc has been derived by \cite{Launhardt+2002}. 
The inferred mass profile has been shown by \cite{Kruijssen+2015} to be well modelled by a power law
$M(r)\propto r^\xi$ with $\xi=2.2$ in the region $45<r/{\rm pc}<115$ (solid body rotation corresponds to $\xi=3$), which matches the radial extent of the gas stream, where the star-forming clouds  are located.
 The fact that $\xi>2$ implies that all the components of the tidal tensor are compressible.
The complete derivation of the equations is given in Appendix \ref{app-tidal}. 
%For sake of clarity, we will omit below the tilde on the variables, keeping in mind that they are all normalized to the Jeans length, Jeans mass and mean freefall timescale
As shown in this appendix, the global collapse condition for an ellipsoidal surdensity ${\rho_R/  \rbar}$ of scale $R$ in the presence of a tidal field becomes
$e^{s_R} \ge e^{s_R^c}$ with:
\begin{equation}
    \label{rho_R_c}
e^{s_R^c}=\frac{2}{I}\frac{1}{\tilde{c}^2}  \left[(1+\mathcal{M}_\star^2\tilde{a}^{2\eta}) -\frac{\mu}{15}\tilde{c}^2\left((\xi-2)+\frac{\tilde{b}^2}{\tilde{c}^2}+\frac{\tilde{a}^2}{q_\phi^2\tilde{c}^2}\right)\right],
\end{equation}
where  $\tilde{a}, \tilde{b}, \tilde{c}$ are the semi axis of the ellipsoidal cloud, with $\tilde{a}$ its longest length,  normalized to the Jeans length, $I$ the normalized deformation tensor, \(\mu=M(r_0)/\bar{\rho}r_0^3\simeq 0.08\) at $r_0=$ 100 pc is the tidal factor, and $q_\phi$ produces a potential flattened in the $z$-direction ($q_\phi=1$ corresponds to a spherically symmetric potential). The best-fitting value to the orbital parameters of the observed clouds yields \(q_\phi=0.63\)\  \citep{Dale+2019}.
The first term in the bracket of eqn.(\ref{rho_R_c}) is the usual HC08 barrier (see eqn.(\ref{eqn-HC})) while the second term is the tidal contribution.
Assuming, for sake of simplicity, that the turbulence-induced perturbations remain nearly spherical
($a=b=c$) the threshold collapse density becomes:
%The initial deformation of the perturbation will a priori not change much the dynamic: it can be seen as a small change of the tidal term. The barrier is much uniformly decreased. 
%To remain within  the range of validity of the HC08 theory, we consider only scales \(R<R_1\) such that \(e^{s_c}>1\) defined by the solution of:
\begin{eqnarray}
    \label{eq_R1}
    &e^{s_{\tilde R}^c}& = {\rho_R^c\over \rbar} = \frac{1+\mathcal{M}_\star^2{\tilde R}^{2\eta}}{{\tilde R}^2}-\mu\kappa, \\
    \Leftrightarrow &{\tilde M}_R^c&\approx \rho_R^c {\tilde R}^3= {\tilde R}(1+\mathcal{M}_\star^2{\tilde R}^{2\eta}) - {\tilde R}^3\mu\kappa, \label{eq-MR}
\end{eqnarray}
where $\kappa=((\xi-2)+1+q_\phi^{-2})/15\simeq 0.22$.
We see that the collapsing barrier is decreased uniformly by the tidal field by a very small factor, $\mu\kappa\approx 0.02$ for the relevant parameter values. The tidal field thus (slightly) favors collapse.

\noindent Although the tidal parameter is very small, it is nevertheless interesting to examine its impact on the slope of the CMF. Indeed, one can imagine that the role of the tides might be increased by more complicated non axisymmetric mass distribution in the CMZ. Since the collapse barrier is decreased compared with the usual one, we need to verify that the condition below is fullfilled:
\begin{equation}
   {\rho_R^c\over \rbar} =  e^{s_{\tilde R}^c} =\frac{1+\mathcal{M}_\star^2R^{2\eta}}{R^2}-\mu\kappa>1.
\end{equation}
%{\bf corresponding to the fact that the density of the turbulence induced perturbation is above the average density.}
This introduces a maximum scale $R_1$ for collapse:
\begin{equation}
    \label{eq_R1}
     {\tilde R}\le  {\tilde R}_1\simeq\left(\frac{\mathcal{M}_\star^2}{1+\mu\kappa}\right)^\frac{1}{2-2\eta}\approx \mathcal{M_\star}^{3.3}.
    %\ \text{and} \ R_2\simeq\left(\frac{(2\eta+1)\mathcal{M}_\star^2}{3\mu\kappa}\right)^\frac{1}{2-2\eta}.
\end{equation}
Furthermore, as seen from eqn.(\ref{eq-MR}), the mass-size relation is no longer monotonic. 
This implies one more condition for collapse and thus another specific scale ${\tilde R}_2$:
\begin{equation}
    \frac{d{\tilde M}_R^c}{d{\tilde R}}\ge 0 \Rightarrow  1+\mathcal{M}_\star^2(2\eta+1){\tilde R}^{2\eta}-3\mu\kappa {\tilde R}^2 \ge 0.
\end{equation}
%This implies another particular scale ${\tilde R}_2$.
 For ${\tilde R}_2\le {\tilde R}_1$, $d{\tilde M}_R^c/d{\tilde R}>0$, ${\tilde M}_R^c$ 
decreases with ${\tilde R}$, the clump will keep being gravitationally unstable, whereas in the opposite case a clump with a density above the threshold value may  eventually become stable again after its initial collapse. Since we are interested in the large scales of the IMF, \({\tilde R}\gg 1\), we can estimate the maximum scale for pursuing collapse, ${\tilde R}_2$, as:
\begin{equation}
    \label{eq_R2}
 {\tilde R}\le {\tilde R}_2\simeq \left(\frac{(2\eta+1)\mathcal{M}_\star^2}{3\mu\kappa}\right)^\frac{1}{2-2\eta}\approx 370\,\mathcal{M_\star}^{3.3}.
\end{equation}
It is easy to verify that, under the present conditions, ${\tilde R}_2/{\tilde R}_1\approx  2/3\mu\kappa\gg 1$. Therefore, if condition (\ref{eq_R1}) is fullfilled, so is condition (\ref{eq_R2}).
A case ${\tilde R}_2<{\tilde R}_1$ would imply $\mu\kappa\gtrsim (2\eta+1)/(2-2\eta)\approx 4$. Under such conditions, tidal effects could stabilize initially gravitationally unstable density fluctuations, modifying drastically the formation of prestellar cores. Given the relevant values of \(\mu\) for the CMZ, however, such a regime does not to occur in this region.

It is not simple to compute the exact dependence of the density upon the critical mass under the combined action of turbulence and tides. However, we can approximate \( {\tilde M}\propto {\tilde R}^\gamma\) with \(\gamma\simeq 2\eta+1 \simeq 2\) when turbulence dominates over tidal effects (see HC08) and \(\gamma\rightarrow 0\) when \({\tilde R}\rightarrow {\tilde R}_2\)
(since $d{\tilde M}_R^c/d{\tilde R}= 0$ for this scale). Thus, the dominant term in the expression of the IMF scales as (eqn.(\ref{eqn-imf})):
\begin{align}
   {\tilde M}{\mathcal N}({\tilde M})= \frac{dN}{d\log {\tilde M}_R} &\propto {\tilde M}_R^{-(\alpha -1)} \\
   &\propto \frac{d\e^{s_R^c}}{d{\tilde R}}\frac{d{\tilde R}}{d{\tilde M}} \\
   &\propto {\tilde M}^{-1-\frac{2}{\gamma}(1-\eta)}.
\end{align}
Note that the first derivative, $d\e^{s_R^c}/d{\tilde R}$, is not impacted by the tidal contribution since the term modifying the usual barrier does not depend on $R$. As $\gamma\to 0$, the slope of the IMF becomes steeper and steeper, as expected since the tidal field favors the collapse, as mentioned above. It is interesting to note, at this stage, that for clump sizes $R\gtrsim 1$ pc, \cite{Tanaka+2020} (their Fig. 3) derive a much steeper mass-size relation, \({\tilde M}\propto {\tilde R}^\delta\), with \(\delta\simeq 2.7\). Assuming that these clumps are approximately in equilibrium, this corresponds to a barrier  \(\rho\propto R^{\delta-3}=R^{-0.3}\). The dominant term in the expression of the IMF thus becomes: ${\tilde M}\mathcal{N}({\tilde M})\propto {\tilde M}^{-3/\delta} \propto {\tilde M}^{-1.1}$, flattening the high mass slope of the IMF but in any case still steeper than the observed slope. As mentioned earlier, however, the aforementioned scales are more characteristic of clouds than cores. For these latter typical scales ($\lesssim 1$ pc), \cite{Tanaka+2020} recover a value \(\delta\simeq 1.6\), in agreement with our above estimate $0\lesssim \gamma \lesssim 2$.

In summary, the tidal field slightly {\it steepens} the slope of the IMF compared with the usual Salpeter value.
%except when using the steep mass-size relation suggested by \cite{Tanaka}. Even in that case, however, the impact is too small to explain the observed high mass slope.
As mentioned previously, the main reason for this small impact of tidal interactions is the high compressibility of the tidal field for the inferred potential of the CMZ.

\section{The Galactic shear}
\label{sec-shear}

Because it is located near  the GC, the CMZ is also subject to a strong shear. The role of the shear has been suggested in many studies to explain the low SFR in the CMZ \citep[e.g.,][and reference therein]{Colling+2018, KrumholzKruijssen2015, Krumholz+2017, Henshaw+2023}. The impact of the shear on the IMF, however, has never been studied in details (see, however, \cite{Dib+2012} for a study on the observational side). 

\noindent This impact is twofold. First, it contributes to the velocity dispersion, second it distorts the shape of the star-forming clumps. We examine both effects in details.

\subsection{The shear as a source of velocity dispersion}
\label{shear_dissipation}

\begin{figure}
    \centering
    \includegraphics[scale=0.6]{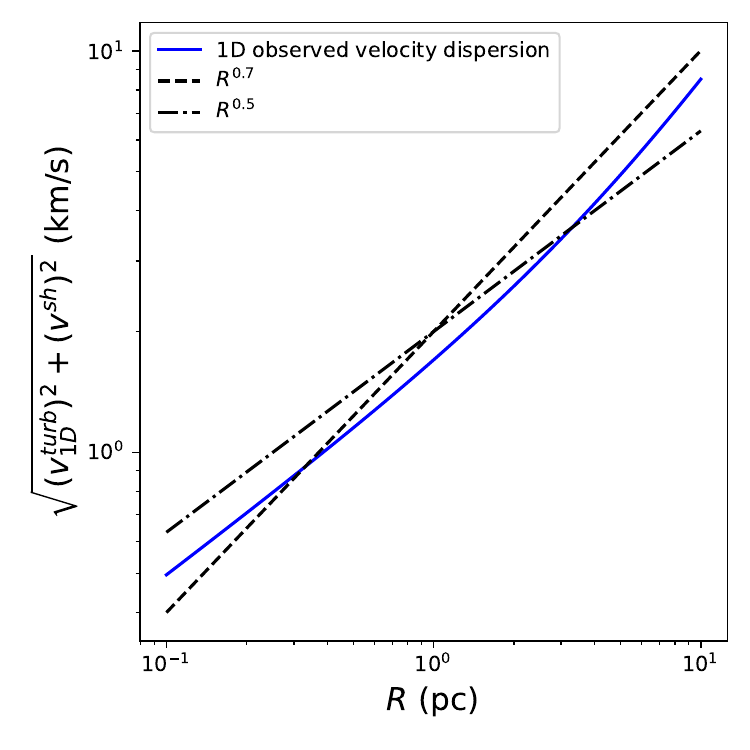}
     \includegraphics[scale=0.6]{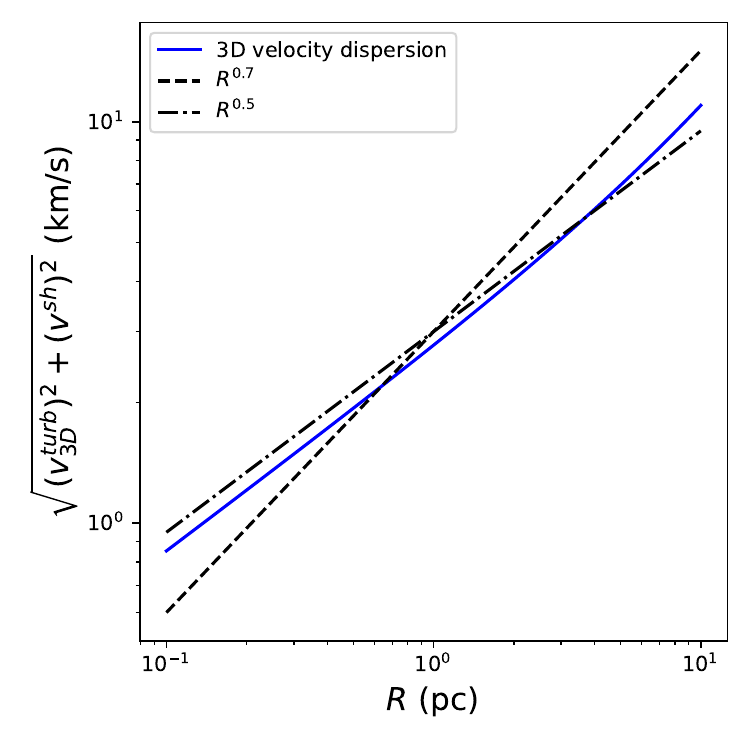}
    \caption{Top: 1D (observed); bottom: 3D total velocity dispersion.}
    \label{velocity}
\end{figure}

The calculations of the impact of the shear on the velocity dispersion are detailed in Appendix \ref{app-shear}.
As mentioned in \S\ref{sec-CMZ}, the CMZ velocity dispersion seems to be well described by the Larson relation, although with an exponent  ($\eta\simeq 0.7$) larger than the usual value ($\eta\simeq 0.5$). This reflects the fact that the  shear gradient contributes to the velocity dispersion, besides  turbulence (see e.g., \cite{Kruijssen+2019a}).
%At the 1 to 100 pc scales, it has been determined by \cite{Krieger+2020} with the tracers CO(3-2) and CO(1-0). 
It must be kept in mind that  observations are done  along a line of sight, so the observed velocity dispersion is the 1D total velocity dispersion:
\begin{equation}
    V_{1D}=\sqrt{(V_{1D}^{shear})^2+(V_{1D}^{turb})^2},
\end{equation}
where (see  eqn.(\ref{eq-shear-vel}))
\begin{equation}
    V^{shear}=\frac{3-\xi}{2}\sqrt{\frac{GM{(r)}}{r^3}}R=0.7\left(\frac{R}{1\text{pc}}\right)\,\kms,
\end{equation}
where \(\xi=2.2\) is the coefficient of the enclosed mass distribution 
$M(r)\propto r^\xi$ determined from the gravitational potential of the CMZ, with  \(M=5\times 10^8 M_\odot\) at \(r=90 \text{pc}\) \citep{Kruijssen+2019a}.
Here, we have used the value of the velocity normalization at 1 pc of \cite{Krieger+2020}, $V_0=2\,\kms$. Indeed, as mentioned previously, this value has been determined in the range 1 to 100 pc whereas the observations of \cite{Tanaka+2020} concern the smaller scales ($\le 1$ pc), where the shear becomes negligible. The turbulent velocity dispersion scales as \(V^{turb}\propto R^\eta_{turb}\) with \(\eta_{turb}\simeq 0.5\) for Burger's like turbulence, while \(V^{shear}\propto R\). The turbulent dispersion velocity at 1 pc is thus given by:
\begin{equation}
    V_{1D}^{turb}(1\text{pc})=\sqrt{V_{1D}^2(1\text{pc})-(V^{shear})^2(1\text{pc})}\simeq 1.6\, \kms.
\end{equation}
Using a value $\eta_{turb}\simeq 0.5$ for the index of turbulence,\begin{equation}
    V_{1D}^{turb}=1.6\, \left(\frac{R}{1\,\text{pc}}\right)^{\eta_{turb}}\,\kms,
\end{equation}
the scale at which the shear induced motions will start to dominate the turbulent ones is given by:
\begin{equation}
    R_{shear}=\left(\frac{V_{1\text{pc}}^{turb}}{V_{1\text{pc}}^{sh}}\right)^{\frac{1}{1-\eta_{turb}}}\text{pc}\simeq 5\,\text{pc}.
\label{rshear}
\end{equation}
This result is consistent with the value of 8 pc above which \cite{Federrath+2016} show that the shear dominates the turbulent motions for the Brick cloud.
These authors, however, supposed in their study that the shear leads to a 3D dispersion velocity whereas the shear yields essentially 1D motions. This assumption increases the role of the shear.
Since at large scale turbulence is isotropic, the turbulent velocity dispersion dominates the total 3D velocity dispersion over a much broader range of scale. Because of the $\sqrt{3}$ correction for $V^{turb}_{3D}$ in eqn.(\ref{rshear}), in reality the shear dominates the turbulent dispersion only for \(R\gtrsim15\) pc. The predicted scalings of the 1D and 3D velocity dispersions are plotted in Figure \ref{velocity}.
\noindent According to eqn.(\ref{scaling_density}), the shear will thus affect the formation of large MCs, for $M\gtrsim 10^5\msol$.
 This is consistent with the study of \cite{Jeffreson+2018}: these authors suggest that while at large ($R\gtrsim 100$ pc) and small ($R\lesssim 45$ pc) galactocentric radii, the cloud lifetime in the CMZ is dominated by galactic shear, cloud evolution in
the zone in-between, which corresponds to the region of compressible tidal field, is dominated by dynamically compressive mechanisms. Consequently, clouds at short and large galactocentric distances are expected to be sheared apart before collapse.
  %, about the mass of the Brick (\(M_{Brick}=7.2\times 10^4 M_\odot \ ; \ L_{Brick}=8 \text{pc}\)).

%As will be shown below, the shear increases the collapsing barrier and thus prevents the formation of such large scale clouds in the CMZ.
Therefore, at the $\lesssim 10$ pc scale, i.e. for the star-forming clouds, the shear plays no or a modest role because the medium is very dense. It might affect indirectly
the mean SFR of the whole CMZ - since fewer clouds are formed, fewer stars are formed - but not the very star formation process, thus the IMF.  Therefore, at these scales, 
we should expect the exponent of the velocity-size relation to be \(\eta\simeq 0.5\), the usual Larson coefficient characteristic of the compressible turbulence cascade, instead of the observed value \(\eta=0.7\) for the determination of the IMF. Indeed, it is worth noting that there is a significant dispersion in the data (see Fig. 4 of \cite{Tanaka+2020}) and a value \(\eta=0.5\) is well compatible for sizes $\lesssim 1$ pc and is probably more physically justified. We will thus use \(\eta= 0.5\) as the fiducial value for the velocity-scale at the core scales in our calculations for the IMF. This well established Burgers value for shock dominated turbulence is confirmed by most observations in molecular clouds (see e.g., review by \citet{HennebelleFalgarone2012}).
It should be noted that \citet{Henshaw+2020} suggest a value \(\eta=0.37\) for the Brick. We have verified that using this value does not affect qualitatively our results, presented in \S\ref{sec-dyn-imf}. Note that these observations rely on one single cloud in the CMZ, with a relatively low large-scale rms Mach number, namely ${\cal M}\simeq 11$ \citep{Federrath+2016}. 
%Furthermore, almost no core is found in the Brick, very likely because of this relatively weak level of turbulence. 
Therefore, the substructure in the Brick observed by  \citet{Henshaw+2020} is unlikely to be representative of the star-forming cloud population in the CMZ.

%The two Larson relations are:
%\begin{equation}
%    \sigma_{1D}(R)=(1.8 \text{km/s})\left(\frac{R}{1\text{pc}}\right)^{0.7} \ \text{and} \ \bar{n}=(3.8\times 10^3 \text{cm}^{-3}) \left(\frac{R}{1\text{pc}}\right)^{-0.3}
%\end{equation}

As shown in appendix \ref{app-shear}, the collapse condition with the shear becomes (see  eqn.(\ref{eq-shear})):
\begin{equation}
    e^{s_{c}}=\frac{2}{I}\frac{1}{\tilde{c}^2}\left[(1+\mathcal{M}_\star^2\tilde{a}^{2\eta})+\left(\frac{3-\xi}{2}\right)^2\frac{\mu}{3}\tilde{c}^2\right],
\end{equation}
where \(\mu=M(r_0)/\bar{\rho}r_0^3\simeq 0.43\). The expression is similar to the one due to the tidal force (eqn.({\ref{eq_R1}})), but, in contrast to this latter, the shear always acts as a {\it support} to star formation, and  it vanishes in case of solid body rotation (\(\xi=3\)), as expected. The reason is that the shear tears apart the gas accumulating in a given region, softening the local gravitational potential and thus allowing more gas to accumulate before collapsing eventually. As in the tidal case, we can estimate the impact of this support on the CMF. Using again \(M\propto R^\gamma\) with \(\gamma\simeq 2\eta+1\) when the shear is negligible compared to the turbulence and \(\gamma\rightarrow 3\) when \(R\gg 1\), the dominant term in IMF scales as:
\begin{equation}
   \frac{dN}{d\log {\tilde M}_R}\propto \frac{d\e^{s_R^c}}{d{\tilde M}}\propto {\tilde M}^{-1-\frac{2}{\gamma}(1-\eta)}.
\end{equation}
As \(\gamma\) varies from \(2\eta+1\approx 2\) to 3, the slope varies from \(-1.25\) to \(-1.2\). The impact on the slope of the IMF due to the shear velocity dispersion is thus very small. 

\noindent It should be mentioned that some studies have suggested that the impact of the shear on the dispersion velocity is comparable to the one due to turbulence \citep{Federrath+2016, Petkova+2023}.
As mentioned previously, however, these studies suppose that the shear leads to a 3D dispersion velocity whereas it should affect essentially 1D motions, yielding a smaller global impact. 

\subsection{The shear as a source of deformation}

%The role of the shear in lowering the star formation rate has been suggested both in the solar neighbourhood \citep{Colling_ImpactGalacticShear2018} and in the CMZ \citep{Li_MeanDensity1122020}. In the picture presented in these papers, 
The second effect of the shear is to distort the star-forming clump. 
%In both studies, however, the background density is much smaller than the one characteristic of the CMZ {\bf CHECK}.  
The deformation between the semi-axis of a clump of size $R$ due to the shear gradient during a typical turbulent timescale, assuming the gas velocity is the shear velocity, can be estimated as (see eqn.(\ref{eq-shear-vel})):
\begin{eqnarray}
    \frac{a}{c}&=&1+2\nabla v_r\tau_{ct} \\
    &=&1+0.6\left(\frac{V_{rms}}{V_0}\right)^{-1}\left(\frac{R}{1\,\text{pc}}\right)^{0.3},
\end{eqnarray}
where $\tau_{ct}={R}/{V^{1D}_{rms}}$ is the typical turbulent crossing time at scale $R$ and where we have taken $V_0=2\,\kms$ and $\eta=0.7$ for the velocity-size relation for the cloud.
This value is comparable  to the one found in \cite{Krieger+2020} and \cite{Tanaka+2020}, between 1.2 and 1.6. 
For a collapsing cloud, however, a more relevant timescale is the free fall time. Replacing $\tau_{ct}$ by $\tau_{ff}$ in the equation above, we get a value $\frac{a}{c}\approx 1.07$ for a 4 pc clouds like the Brick.

\noindent The correction due to the shear deformation upon the collapsing barrier can be estimated by considering the integral of the deformation tensor \citep{Chandra87},
$I(a/c=1.6, b/c=1)\simeq 1.9$ and $I(a/c=1.07, b/c=1)\simeq 1.09$. The collapsing barrier is thus slightly increased  because of the stabilising effect of the ellipsoidal deformation.

Therefore, although the shear can have a substantial stabilising effect for large scale, non-collapsing clouds, it is negligible for small-scale collapsing ones.

\bigskip

%In summary, a correction of the slope of the CMF is found for the mass-size and velocity-size relations derived \cite{Krieger+2020}  but not for those found in \cite{Tanaka+2020}. The correction is particularly noticeable when considering the ellipsoidal collapse condition \cite{DumondChabrier2024}. However, this impact of the shear on the slope of the IMF remains too small to explain the observed flattening of this latter.

%The pic is however translated to the large masses by a factor 3 or 4 compared to the IMF found by \cite{Hosek+2019}. As it is  translated to smaller masses for the Tanaka's relation, we can expect than intermediate meaningful scaling factors will lead to the translation of the pic to lower masses. Moreover, as we consider here the CMF and compare with the IMF, it is expected that the CMF is translated to higher mass with respect to the IMF due to a uniform core to star efficiency factor (3 if the usual one is relevant in the CMZ).

%The role of the shear in lowering the star formation rate has been suggested both in the solar neighbourhood \citep{Colling_ImpactGalacticShear2018} and in the CMZ \citep{Li_MeanDensity1122020}. In the picture presented in these papers, 
%In both studies, however, the background density is much smaller than the one characteristic of the CMZ {\bf CHECK}.  

\bigskip

The two above sections show that both the shear and the tides {\it slightly} affect the IMF slope of star-forming clumps in the CMZ (either as a source of deformation or velocity dispersion) within about the same order of magnitude, although with opposite contributions, yielding a global negligible effect. For sake of completeness, we have examined in App. \ref{app-tide-shear} the cumulative effect of tides and shear.  In order to maximize the effects, we have considered ellipsoidal density fluctuations. The global result is that the collapsing barrier is found to be lower than the usual HC one. This was intuitively expected since while the shear acts only in 1 dimension, the tides act in the 3 directions and thus dominate the global impact. However, the final result remains unchanged as the global effect is found to be too small to explain the shallow IMF. 
It is worth pointing that the results strongly depend on the chosen parameters at the cloud scale,
\(b\), \(V_0\) and \(\nbar\). The impact on the CMF is maximized  for a high velocity dispersion (typically 10 $\kms$) and low density \(\nbar=2\times 10^3\,\cc\). Such densities, however, seem to be inconsistent with the star-forming main cloud conditions observed in the CMZ.

\section{The dynamics of star formation in the CMZ}
\label{sec-dynamics}

\subsection{The Mass Function}
\label{sec-dyn-imf}
%This suggests that large scale turbulence is injected sporadically instead of being constantly sustained like in the Milky Way disk. In our present understanding of the gravoturbulent scenario of star formation, this has a crucial impact, as will be seen. Indeed, since the largest free-fall time of a massive star, $\tff\simeq (3\pi/32G\rho(R))^{1/2}$ is shorter than 100 Myr, this implies that all the stars, thus the IMF, are formed during one single turbulent episode. {\bf crossing time ?} In other words, the cycle of star formation occurs during one turbulent dissipation timescale, before a new bursts occurs eventually. The origin of the SFR variations, however, variable bar-driven inflow rate or stellar feedback. Numerical simulations of this latter process, for instance, still yield diverse results.
%\subsection{The core mass function}

There are observational suggestions that star formation in the CMZ does not occur continuously but through episodic star bursts \citep{Yusef-Zadeh+2009,Kruijssen+2014,Henshaw+2023}. This is supported by numerical simulations which suggest bursts lasting about
$\sim$5-10 Myr and separated by about $\sim20$-40 Myr, about the time for the clouds to migrate inwards to the gravitationally unstable region at $R\sim100$ pc \citep{KrumholzKruijssen2015,Krumholz+2017}. A process also found for other extragalactic nuclei (see review by \citet{Henshaw+2023}). The fact that the CMZ would be presently between two star-bursts provides a plausible explanation for its particularly low SFR \cite[see e.g.,][]{KrumholzKruijssen2015}.
%The fact that the most massive stars observed in the CMZ are about 30 \(M_\odot\) \citep{Hosek+2019} corroborates this scenario since 40 Myr is about the lifetime of a 30 \(M_\odot\) star. {\it This also naturally explains the severe drop in the slope of the IMF observed at this mass  for the Arches} \citep{Hosek+2019}. More massive stars have already exploded as supernovae. 

As shown, e.g., by \cite{Jeffreson+2018}, clouds are expected to collapse and form stars  at radii from $\sim$45 to 100 pc,
where shear is reduced and gas accumulates, while at short and large galactocentric distances the clouds are expected to be sheared apart before collapse and star formation occur, yielding low star-formation efficiency  \cite[see e.g.,][]{KrumholzKruijssen2015,Jeffreson+2018}.
% while at large ($R\gtrsim 100$ pc) and small ($R\lesssim 45$ pc) galactrocentric radii, the cloud lifetime in the CMZ is dominated by galactic shear
%, cloud evolution in the zone in-between, which corresponds to the region of compressible tidal field, is dominated by dynamically compressive mechanisms. Consequently, clouds at short and large galactocentric distances are expected to be sheared apart before collapse and star formation occur, yielding low star-formation efficiency, while 
Because  of their high densities, these clouds have short lifetimes, of $\sim$0.3-4 Myr  \citep{Kruijssen+2015,Jeffreson+2018}. Interestingly enough, this corresponds to about one turbulence crossing time,
$\tau_{ct}\simeq L_c/V_{rms}^{1D}$, for the clouds of interest, with $L_c\sim1$-10 pc. Furthermore, given their short typical lifetimes compared with the timescale between 2 starburst episodes (see above), the clouds  have experienced only one such episode before collapsing. In any case, the fact \(\tau_{cloud}<\tau_{ct}\) implies that the cloud experiences only one turbulent episode during its lifetime.
According to the Hennebelle-Chabrier time-dependent theory \citep{HennebelleChabrier2011,HennebelleChabrier2013}, the crossing time at the cloud scale is the typical time that is necessary for the density field 
generated by large-scale turbulence to be significantly modified, triggering a new set of density fluctuations, statistically independent of the former one. 
In strongly magnetized flows, as in the present case, the time required to rejuvenate a self-gravitating structure is even longer.
%According to the HC or H core formation theory, every episode of turbulent flow across the cloud will generate overdensities at scale $R$ of mass $M_R$ of which the gravitationally instable ones will collapse and form eventually a prestellar core. Therefore, the number of prestellar core fluctuations depends of their scale, with $\sim \tau_R/\tau_0$ collapsing fluctuations of scale $R$ being created during the lifetime $\tau_0$ of the cloud.
Therefore,  star formation within clouds in the CMZ should occur rapidly, throughout {\it one single episode of large-scale injection of turbulence}.

In that case, the static, time-independent approach of the HC theory should be used instead of the time-dependent one. Indeed, in this latter case, small scale density fluctuations
can keep collapsing during the collapse of large ones (which have a longer free-fall timescale) because turbulence
is constantly replenished at the the cloud scale during the cloud lifetime ($\tau_{ct}<\tau^0_{ff}$). In contrast, for the CMZ conditions, the above estimate shows that the cloud has collapsed before the first episode of turbulence has had time to dissipate and a new one will be generated at large scales.
 
\begin{figure}
    \centering
    \includegraphics[scale=0.6]{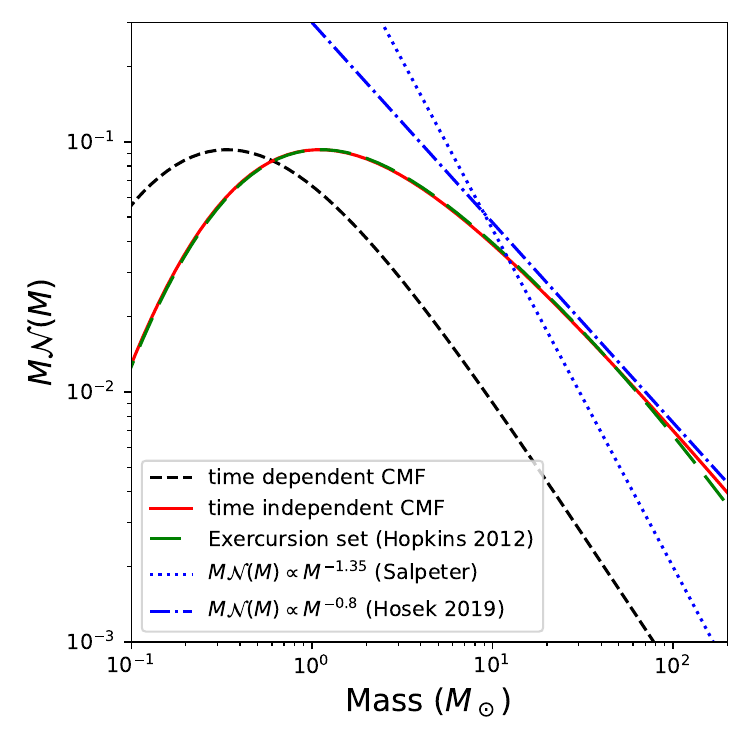}
     \includegraphics[scale=0.6]{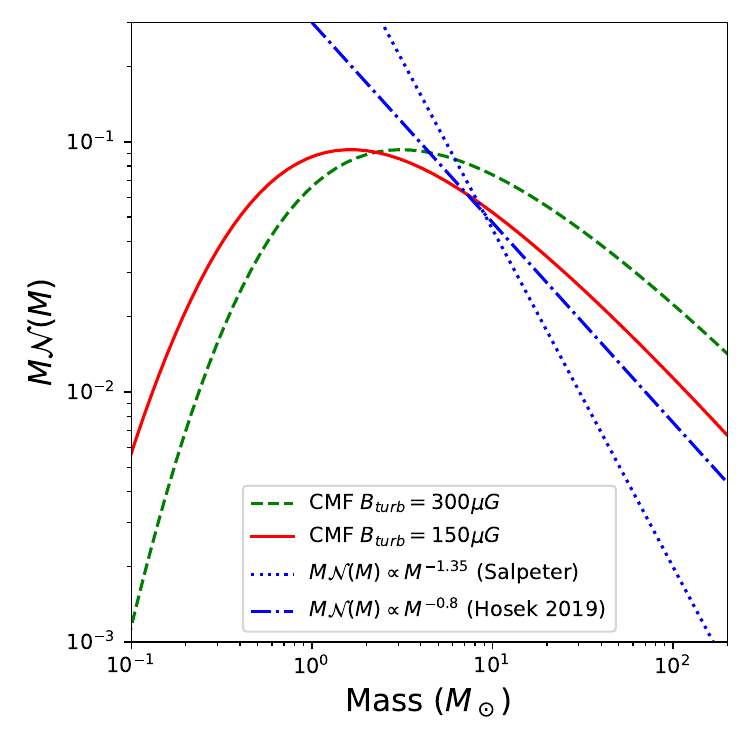}
    \caption{Top: Initial mass function $dN/d\log M$ calculated with the time-dependent (HC13, short dashed line) and time-independent (HC08, solid line) Hennebelle-Chabrier theory for a magnetized cloud of size \(L_c=10\) pc with no impact of the magnetic field upon the PDF ($B\propto n^0$). Long dashed line: time independent IMF calculated with the \cite{Hopkins2012} excursion set formalism. Dotted line: Salpeter IMF; dash-dot line: IMF observed for the Arches \cite{Hosek+2019}. Bottom: IMF calculated with the time-independent (HC08) theory for 2 representative values of the turbulent magnetic field from \cite{Lu+2023} taking into account the modification of the variance of the PDF (eqn.(\ref{variance})), which corresponds to $\beta=0.25$ and 1, respectively. This supposes a field-density dependence $B\propto n^{1/2}$ and we have taken the velocity normalization of \cite{Tanaka+2020}.}
        \label{fig-imf}
\end{figure}

 As shown in HC13, the time-independent description of the IMF yields a shallower IMF than the time-dependent one, by a factor $\Delta \alpha = (\eta-1)/(2\eta+1)$ (see their eqns.(24), (25)). This is illustrated in 
 the top panel of Fig. \ref{fig-imf}. While the time-dependent derivation yields a too steep high-mass tail for the IMF, the time-independent formulation properly reproduces the observed value. We stress the fact that there is no adjustable parameter in this theory. It is just applied in the present context to the characteristic properties of the CMZ star-forming clouds (\S\ref{sec-CMZ}). In order to verify this result, we have carried out the same type of calculations with \cite{Hopkins2012} excursion set formalism (long dashed line in Fig. \ref{fig-imf}). As seen in the figure, the result is confirmed.
 
\noindent It is worth noting that, as seen in the figure, the time
 dependent calculation not only fails reproducing the correct value for the Arches, but it does not even recover  the Salpeter value. This seems to be at odds with figures 1 and 2 of HC13 that show that, for standard MW clouds, the time-dependent theory yields the correct slope.
 The key point is that while usual Larson MW conditions correspond to ${\mathcal M}_\star\simeq {\sqrt 2}$, the present ones give ${\mathcal M}_\star\simeq 10$ for the velocity dispersion appropriate to the core scales \citep{Tanaka+2020} (see \S\ref{sec-velocity-size}). As shown in HC08 (their Fig. 1), the larger  ${\mathcal M}_\star$ the shallower the high-mass slope, a consequence of the more vigorous turbulence cascade. Such levels of turbulence are reached in massive early type galaxies, even though the proper slope of the IMF is recovered with the time-dependent HC theory \citep{CHC14}. These galaxies, however, are at least 10 times denser than in the present case (see Table 1 of \cite{CHC14}), increasing the role of gravity to counteract the one of turbulence in the collapse process. This confirms the fact that, for typical density-turbulence conditions in CMZ star-forming clouds, an ongoing injection process of large scale turbulence can not produce the observed high-mass slope, in contrast to the 'one single turbulence injection episode' scenario.
  
%We see that our hypothesis of 'one single turbulence injection episode' for the formation of prestellar core condensations in the star-forming clouds of the CMZ provides a plausible explanation to explain its shallow high-mass tail.
For sake of completeness, we also show in the bottom panel of figure \ref{fig-imf} the IMF obtained in the time-independent case if the PDF of the cloud is modified by the strong magnetic field for the (unlikely) case $B\propto {\nbar}^{1/2}$, as discussed in \S\ref{sec-magnetic}. We have used the more appropriated velocity dispersion value of \cite{Tanaka+2020} for the cores (see discussion in \S\ref{sec-pdf}). In that case, the high mass slope of the CMF is flattened and becomes consistent with the observed value. However, such a field-density amplification is unlikely to occur for CMZ cloud densities. Indeed, the magnetic field is usually found to saturate at densities $\lesssim 10^4\,\cc$ \cite[e.g.][]{HuLazarian2023}. As discussed in \S\ref{sec-magnetic}, in that case the magnetic field plays no role on the gas PDF.

%Note that if we pick a value $b=0.2$ for the turbulence driving parameter, as suggested by \cite{Federrath+2016}, meaning that the turbulence is essentially solenoidal, the slope gets even shallower.

As shown in HC08, the transition from the large scale turbulence dominated regime to the small scale thermal one corresponds to a mass ${\tilde M}\simeq 2({\mathcal M}_\star)^{-1/\eta} \approx 0.02$ under the present conditions against $\sim0.8\,\msol$ for usual cloud conditions (see eqn. (45) of HC08). Therefore, whereas for usual MW conditions, the peak of the IMF ($d{\mathcal N}/d{\tilde M}=0$) occurs in the thermal regime, under the present conditions it occurs in the {\it turbulence dominated} one, as shown in Fig. \ref{fig-imf}. 
From the physical point of view, this means that under CMZ cloud conditions, the vigorous turbulence can generate and prevent collapsing a larger number of density fluctuations than under usual conditions. Note also that the warmer gas provides a larger thermal support.  We stress again that, as explained in detail in \cite{ChabrierHennebelle2011}, the role of turbulence in the HC theory of the IMF should {\it not} be considered in a static (pressure-like) sense because turbulence has already dissipated by the time the prestellar core is formed.
Turbulence in the HC sense must be understood as a dynamical, statistical process, turbulence generating {\it at the very initial stages} a field of density fluctuations and sweeping away the ones not dense enough to have a chance to collapse
under the action of gravity, allowing more gas to accumulate locally before becoming gravitationally unstable. By the time a dense enough fluctuation has started to collapse into a prestellar core, the turbulence at the core scale has dissipated and has become transonic or subsonic. What matters in this formalism is thus the  turbulence {\it rms velocity} rather than pressure. 
%However, this is purely coincidental, as the sonic length does not enter explicitly into the theory.

\subsection{The source of turbulence}
\label{sec-turb}

Various sources of turbulence driving in the CMZ have been examined in  \citet{Kruijssen+2014} and  \citet{Henshaw+2023}. Among different processes these authors found that bar inflow could explain the observed turbulent energy. This is confirmed in a study by  \citet{SormaniBarnes2019} who find that gas inflow is a promising candidate for driving the turbulence in the CMZ. It is indeed well known that the bar transports matter and energy radially from the Galactic disk to the CMZ (see e.g.,  \citet{Portail+2017,KrumholzKruijssen2015}). Mass transport is a natural consequence of the disc instability as the strongly non-axisymetric structure of the Galactic bar exert torques that tend to drive angular momentum out and thus yield inward mass transport by angular momentum conservation. When mass is transported inwards through the disk and down the overall potential well, part of the gravitational energy gain can be converted into turbulence (see e.g., \citet{Wada+2002}). Shear stress, notably near the boundary between the (inner) gravitationally dominated and (outer) shear dominated region of the CMZ, will also contribute to turbulent driving at large scale. On the other hand, estimates of energy injected by supernovae in the CMZ, after a first generation of stars has been formed, are also found potentially to contribute significantly to turbulence in the CMZ \citep{Henshaw+2023}. This list, of course, is not exhaustive, and processes like acoustic or MRI disc instabilities might also contribute to the turbulence driving, even though it seems difficult for these processes alone to sustain the observed high levels of turbulence \citep{Henshaw+2023}. 
%Star bursts and SN explosions will also contribute to turbulence injection in the CMZ, but  are probably not the main drivers. 
All these processes, at different levels, constantly drive high levels of velocity dispersion in the CMZ. This provides the seed for the generation of density fluctuations, once enough gas has accumulated again in the potential dominated region, after the left over gas due to the previous star formation episode has been dispersed. This will provide the turbulent dense gas reservoir necessary to generate a new field of density fluctuations, eventually yielding a new star burst episode. Interestingly enough, one note that 30-40 Myr, the expected duration of the quiescent phase, is about the lifetime of a 8 $\msol$ star, the smallest mass for SNII progenitors, before it explodes as a SN. During all this period, supernovae explosions will at the very least slow down the inward gas inflow. This provides a support, besides the aforementioned inward gas migration argument, for the duration of the quiescent phase. Furthermore, as mentioned by \citet{KrumholzKruijssen2015}, ‘quiescent’ does not necessarily mean that the gas is completely depleted, but rather that it is driven out of a self-gravitating state. 

\subsection{The star formation rate}
\label{sec-sfr}

The fact that star formation occurs within one single turbulent  episode will also reduce the SFR. Indeed, once a fluctuation of scale $R$ will have collapsed, it will
not be rejuvenated by a new episode of turbulence cascade dissipation.
In that case the derivation of the SFR per free-fall time yields  (see HC13):
%This corresponds to setting $\tau^0_{ff}/\phi_t\tau_{R,ff}=1$ in eqn.(45) of \cite{HennebelleChabrier2011}, where $\tau_{R,ff}$ is the free-fall time at scale $R$ and $\phi_t\approx$ 2-3. This yields
\begin{align}
    \text{SFR}_{ff}^0 &= \int_{0}^{M_{cut}}\frac{M\mathcal{N}(M)}{\bar{\rho}}dM \\
    &=\int_{\tilde{\rho}_{cut}}^{+\infty} \mathcal{P}(\tilde{\rho})d\tilde{\rho} \\
    &= \frac{1}{2} \left(\text{erf}\left(\frac{\sigma^2-2 \log (\tilde{\rho}_{cut})}{2^{3/2} \sigma }\right)+1\right),
    \label{eq_SFR_no_time}
\end{align}
where \(\tilde{\rho}=\rho/\bar{\rho}\) and we assume that the probability distribution depends weakly on  scale \(R\). As in HC13, we take  \(y_{cut}={\tilde{R}_{cut}}/{\tilde{L}_c}\simeq 0.1-0.3\), which corresponds to 
\begin{equation}
    \tilde{\rho}_{cut}=\frac{1+\mathcal{M}_\star^2\tilde{R}_{cut}^{2\eta}}{\tilde{R}_c^2} \simeq 4 - 14.
\end{equation}

Compared with eqn.(45) of \cite{HennebelleChabrier2011}, we notice the absence of the prefactor $e^{(3\sigma^2/8)}$ which stems from the disappearance of the density free-fall timescale ${\tilde \rho}^{1/2}$ in the integral.

\begin{figure}
    \centering
    \includegraphics[scale=0.6]{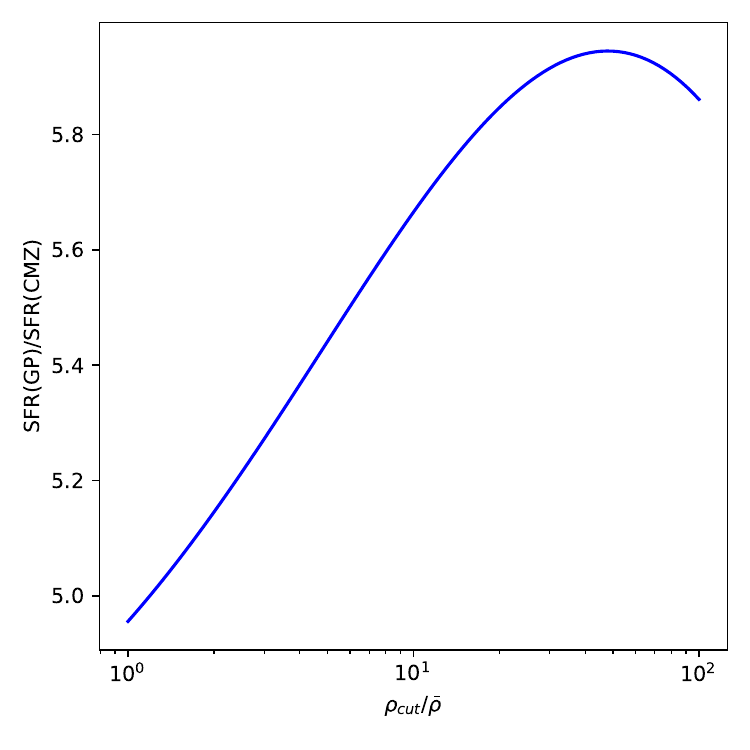}
    \caption{Ratio of the SFR in the galactic plane (GP) and in the CMZ from the HC13 model as a function of \(\tilde{\rho}_{cut}\). The SFR in the CMZ is calculated from the time independent model (eqn. \ref{eq_SFR_no_time}) with parameters characteristic of the CMZ. The SFR in the GP is calculated from the time dependent model from HC13 with parameters characteristic of the GP.}
    \label{SFR_comp}
\end{figure}

Figure \ref{SFR_comp} compares the SFR predicted by the usual time-dependent HC theory for fiducial MC conditions (\(\mathcal{M}=15, b=0.5 \Rightarrow \sigma\simeq 2\)) with the SFR typical of CMZ conditions (\(\mathcal{M}=100, b=0.3 \Rightarrow \sigma\simeq 2.6\)). We see that assuming that star formation in
CMZ MCs occurs throughout one single episode of turbulence injection at the cloud scale reduces the SFR by about a factor 5 to 6. As shown in figure \ref{SFR_comp}, this ratio depends very weakly on the choice of \(y_{cut}\).

A point worth mentioning is that feedback has often been suggested as an explanation to reduce the SFR in the CMZ. At least dynamical feedback, however, is already implicitly taken into account in the stronger than usual velocity dispersion and in the modified index $n$ of turbulence through the Larson parameter $\eta$. Indeed, as shown in HC08 (their eqn.(24)) both indexes are related through the relation $\eta=(n-3)/2$. The observed value $\eta=0.7$ thus implies an index of turbulence $n=4.4$, larger than the Burgers value. It seems thus reasonable to suggest that the shear and the dynamical feedback are the reasons for such a modification of the velocity dispersion relation in the CMZ. Distinguishing between the turbulence and shear contributions, however, is generally not done in the observational analysis. 

\noindent Some simulations have pointed out the shear for lowering the SFR  both in the solar neighbourhood \citep{Colling+2018} and in the CMZ \citep{LiZhang2020,Dale+2019, Emsellem+2015} even though \cite{Dib+2012} show that the impact of the shear upon the SFR is not relevant far from the GC. This seems to be in contradiction with the analysis carried out in \S\ref{sec-shear}.

These studies, however, consider a
much less dense medium (or much larger clouds) than the one characteristic of the CMZ, although with a strong shear. As shown in \S\ref{sec-shear}, the more diluted the medium the greater the impact of the shear. 
Similarly, \cite{Federrath+2016} have suggested that turbulent velocity dispersion, density, and forcing parameter as the ones observed in the Brick  cloud and used in \cite{Kruijssen+2019a} lead to the right SFR. However, as mentioned in \S\ref{sec-shear}, these authors assume in their analysis that the shear acts in the 3 directions. Taking a more reasonable assumption of a 1D action of the shear, we have shown that its effect is much more modest in the region of the star-forming clouds.

\section{Conclusion}
\label{conclusion}

In this paper, we have examined various physical processes that may explain the shallow high-mass slope of the IMF, as well as the low SFR in star-forming MCs in the CMZ. 
We show that neither the strong tidal field nor the Galactic shear experienced by the CMZ can explain these unusual properties, {\it in the molecular gas region where star-forming clouds form}. Both effects
have a negligible impact in this region. Moreover, interestingly enough, shear and tidal interactions have opposite effects. While tides lower the collapsing barrier, promoting the collapse of overdense fluctuations, the shear increases the barrier, bringing support against collapse. Both effects are similar and nearly compensate. 
Similarly, we show that the intense magnetic field in the CMZ provides a negligible pressure support and does not modify the PDF of the turbulent gas flow in the clouds, {\it except if it does not saturate and keeps being amplified 
at these high densities}. An hypothesis which lacks a robust physical explanation and is thus unlikely. 

 In contrast, we show that, contrarily to the case of  MCs in the Galactic disk, clouds in the CMZ experience only one single episode of turbulence injection at large scale, most likely due to bar instabilities and radial mass transport (see \S\ref{sec-turb}).
Indeed, their rather short lifetime, due to their high mean densities, is similar to one typical turbulence crossing time at the injection (i.e. cloud) scale. Consequently, according to the Hennebelle-Chabrier theory of star formation, within this 'single turbulence episode' scenario, the cloud experiences one single field of turbulence induced density fluctuations, leading eventually to gravitationally unstable prestellar cores. New generations of  large scale turbulence induced density fluctuations have no time to occur. As shown in \citet{HennebelleChabrier2013}, for a given cloud mass, this yields a flatter IMF than usual, leading to the correct observed value for the characteristics of the CMZ star-forming clouds. 
  We also suggest that the severe steepening observed in the IMF of the Arches, with an age of $\sim$2-4 Myr, above $\sim40\,\msol$ \citep{Hosek+2019} can be explained by the maximum lifetime of these stars before they explode as supernovae. More massive stars have already exploded as supernovae.

 Furthermore, this single large scale turbulence event within the cloud lifetime yields by itself a 5 to 6 lower SFR than under usual MW cloud conditions, again in agreement with the observed values. Therefore, we conclude that  our 'single large scale turbulence injection' scenario for CMZ star-forming clouds provides a plausible, consistent explanation for both the observed shallow high-mass slope and low SFR in the CMZ.

%%%%%%%%%%%%%%%%%%%%%%%%%%%%%%%%%%%%%%

\appendix

\section{Tidal tensor and virial collapse condition}
\label{app-tidal}

We consider a MC whose centre of mass \(m_0\) is at a distance \(r_0\) from the GC. We also consider a mass \(m_1\ll m_0\) located in the MC at the position \(\vec{r}\) from its  mass centre and at \(\vec{r}_1=\vec{r}_0+\vec{r}\) from the GC. The mass in the CMZ creates an axisymmetric gravitational field \(\phi\). The tidal force on \(m_1\) is the difference between the force exerted on \(m_1\) and the one exerted on the centre of mass of the system \(\{m_0 + m_1\}\):
\begin{align}
    \vec{F}_t 
    %=\vec{F}_1-m_1\frac{\vec{F}_1}{m_0+m_1} 
    =-m_1\left[\vec{\nabla}\phi(\vec{r}_0+\vec{r})-\vec{\nabla}\phi(\vec{r}_0)\right].
\end{align} 
As the mass of the cloud is very small compared with the one generating the tidal field, we neglect its gravitational field. A Taylor expansion in the cartesian frame \(\vec{e}_x, \vec{e}_y, \vec{e}_z\) (as in \cite{Dale+2019}), \(\vec{e}_x \parallel \vec{r}_0\), yields:
\begin{equation}
    \label{tidal_force}
    F_t^j(\vec{r})=-m_1r_k\frac{\partial\phi}{\partial x_k\partial x_j}(\vec{r}).
\end{equation}
The tidal term involved in the Virial theorem is given by:
\begin{equation}
    T_{ii}=\int\rho\vec{f}_t^i r_idV\simeq-\int\rho\frac{\partial\phi}{\partial r_i^2}r_i^2dV,
\end{equation}
where \(r_i=x, y\) or \(z\) and \(\vec{f}_t = \vec{F}_t/m_1\) is the tidal force by unit of mass. Note that in the last equality, we neglected the non-diagonal terms of eq. \ref{tidal_force} because they correspond to negligible corrections. The system of coordinate used here is summarized in figure \ref{Geometry}. In the cylindrical frame associated to the Galactic Center (GC), the radial gravitational field is given by:
\begin{equation}
    \vec{G}(r)=-\frac{GM(r)}{r^2}\vec{e}\,r \Leftrightarrow \phi(r)=\frac{GM(r)}{(\xi-1)r},
\end{equation}
where \(M(r)\propto r^\xi\) is the mass enclosed within a radius $r$, with \(\xi=2.2\) the value determined from the gravitational potential of the CMZ \citep{Launhardt+2002,Kruijssen+2015} (solid body rotation corresponds to $\xi=3$). 
Using \(dr/dy=y/r_0\) and considering the flattening of the gravitational field along the $z$ axis \citep{Dale+2019}, we deduce the three diagonal components of the tensor:
\begin{equation}
    \frac{\partial^2\phi}{\partial x^2}=(\xi-2)\frac{GM(r)}{r^3} \ ; \ \frac{\partial^2\phi}{\partial y^2}=\frac{GM(r)}{r^3} \ ; \ \frac{\partial^2\phi}{\partial z^2}=\frac{GM(r)}{q_\phi^2r^3},
\end{equation}
where    $q_\phi$ produces a potential flattened in the $z$-direction.
As \(\xi>2\), all components of the tidal tensor are compressive. Doing the usual change of variable between cartesian and spherical coordinates for an ellipsoid yields for the components of the tidal tensor:
\begin{align}
    T_{xx}&=\frac{2-\xi}{5}M_cc^2\frac{GM(r_0)}{r_0^3},\\ 
    T_{yy}&=-\frac{1}{5}M_cb^2\frac{GM(r_0)}{r_0^3}, \\ 
    T_{zz}&=-\frac{1}{5}M_cc^2\frac{GM(r_0)}{q_\phi^2r_0^3},
\end{align}
where \(M_c\) is the total mass of the studied molecular cloud and \(a, b\) and \(c\) are the three lengths of the ellipsoid. 

An equation of motion can be written for each axis:
\begin{equation}
    \frac{1}{2}\frac{\partial^2 I_{ii}}{\partial t^2}=PV+2 K_{ii}+W_{ii}+T_{ii},
\end{equation}
where \(I_{ii}, K_{ii},W_{ii}\) and \(T_{ii}\) denote respectively the inertial, kinetic, gravitational and tidal tensors. Denoting \(\tilde{x}\) the quantities normalized to the Jeans length, \(\lambda_J=C_s/\sqrt{\bar{\rho}G}\), Jeans mass, \(M_J=4\pi/3\bar{\rho}\lambda_J^3\), and mean free-fall time \(\tau=1/\sqrt{G\bar{\rho}}\), one gets after calculations:
\begin{align}
    \label{Virial_eq_tides}
    \frac{1}{2}\frac{\partial^2 \tilde{c}^2}{\partial\tilde{t}^2} &=\frac{5}{2}(1+\mathcal{M}_\star^2 \tilde{a}^{2\eta})-\pi\alpha_1 \tilde{c}^2\e^s-\frac{\xi-2}{2}\tilde{c}^2\mu, \nonumber \\
    \frac{1}{2}\frac{\partial^2 \tilde{b}^2}{\partial\tilde{t}^2} &=\frac{5}{2}(1+\mathcal{M}_\star^2 \tilde{a}^{2\eta})-\pi\alpha_2 \tilde{c}^2\e^s-\frac{1}{2}\tilde{b}^2\mu, \nonumber \\
    \frac{1}{2}\frac{\partial^2 \tilde{a}^2}{\partial\tilde{t}^2} &=\frac{5}{2}(1+\mathcal{M}_\star^2 \tilde{a}^{2\eta})-\pi\alpha_3 \tilde{c}^2\e^s-\frac{1}{2q_\phi^2}\tilde{a}^2\mu,
\end{align}
with \(\mu=M(r_0)/\bar{\rho}r_0^3\simeq 0.08\) and \(q_\phi=0.63\) \citep{Dale+2019}. 
%In these equations, the evolution of the turbulence "support" might be debated. Does it evolve during the cloud evolution following the corrected Larson relation (increases when the cloud dissipates and decreases when it collapses) or does it stay constant. There is probably
For sake of simplicity, we neglect the possible modification of turbulence by accretion on or dissipation of the cloud and thus replace \(a^{2\eta}\) by \(a(t=0)^{2\eta}\), where \(a\) is the longest length of the cloud.

\noindent Following the Hennebelle-Chabrier formalism, the global collapse condition is obtained be summing these three equations:
\begin{equation}
    \label{rho_R_c}
    e^{s_{c}}={\rho^{c}_R\over \rbar}={\frac{2}{I} \frac{1}{\tilde{c}^2}} \left[(1+\mathcal{M}_\star^2\tilde{a}^{2\eta})-\frac{\mu}{15}\tilde{c}^2\left((\xi-2)+\frac{\tilde{b}^2}{\tilde{c}^2}+\frac{\tilde{a}^2}{q_\phi^2\tilde{c}^2}\right)\right],
\end{equation}
where $I(a,b,c)=\int_0^\infty\frac{du}{\sqrt{(1+\left(\frac{c}{a}\right)^2u)(1+\left(\frac{c}{b}\right)^2u)(1+u)}}$ \citep{Chandra87}.
The first term in the bracket is the usual HC08 barrier while the second term is the tidal contribution.

 \section{Shear}
\label{app-shear}

\begin{figure}
    \centering
    \includegraphics[scale=0.6]{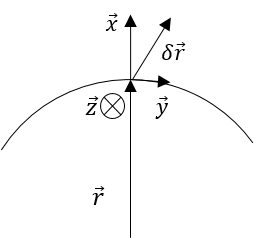}
    \caption{Reference frame for the calculation of the tides and the shear}
    \label{Geometry}
\end{figure}

We consider a rotating frame of angular velocity \(\Omega\) associated to the centre of mass of a cloud located at a distance \(r\) from the GC. Using usual transformations for time derivatives between the rotation and inertial frames (at the GC), we deduce the relation at any point of the cloud:
\begin{equation}
    \vec{v}_r(\vec{r}+\delta \vec{r})=\vec{v}_i(\vec{r}+\delta \vec{r})-\vec{\Omega}(r)\times(\vec{r}+\delta\vec{r}),
\end{equation}
where \(\delta\vec{r}=x\,\vec{e}_x+y\,\vec{e}_y\) in the rotating basis, \(\vec{\Omega}=\Omega\,\vec{e}_z\) and \(\vec{v}_i=\vec{\Omega}(\vec{r}+\delta\vec{r})\times (\vec{r}+\delta\vec{r})\). Thus:
\begin{align}
    \vec{v}_r =(\vec{r}+\delta\vec{r})\times\left(\vec{\Omega}(\vec{r}+\delta\vec{r})-\Omega(\vec{r})\right) 
    =(\vec{r}+\delta\vec{r})\times\left[(\delta\vec{r}\cdot\vec{\nabla})\vec{\Omega}(\vec{r})\right].
\end{align}
We define the axis orientation illustrated in figure \ref{Geometry}. 
We thus have:
\begin{equation}
    r_1=|\vec{r}+\delta \vec{r}|=r\sqrt{1+\frac{x^2+y^2}{r^2}+2\frac{x}{r}}.
\end{equation}

At  first order, we deduce \(r_1=r+x\). Then, using this expression, we have 
\begin{align}
    \vec{v}_r &=-(\vec{r}+\delta\vec{r})\times x\frac{d\Omega}{dr}\vec{e}_z =-rx\frac{d\Omega}{dr}\vec{e}_y.
\end{align}
The minus sign comes from the fact that \(\vec{\Omega}\) is oriented in the direction opposite  to \(\vec{e}_z\). To compare with the conditions used in simulations \citep{Dale+2019, Kruijssen+2019a}, we give the numerical value of the velocity gradient:
\begin{eqnarray}
    \nabla v_r&=&r\frac{d\Omega}{dr}=\frac{3-\xi}{2}\sqrt{\frac{GM(r)}{r^3}}\nonumber\\
    &\simeq& 1\times10^{-14}\, \text{s}^{-1}\simeq 0.7 \,\text{Myr}^{-1}\,\simeq 700\,\kms \text{kpc}^{-1},
\end{eqnarray}
where \(\xi=2.2\) is the coefficient of the mass distribution  $M(r)\propto r^\xi$ determined from the gravitational potential of the CMZ, with  $M=5\times 10^8 M_\odot$ at $r=90 \text{pc}$ \citep{Kruijssen+2019a}.
This value is larger than the one used by \cite{Colling+2018}: this is due to the fact that in these simulations, they are far from the GC.
At first order, the velocity dispersion only exists along the \(y\) axis, but at  second order there is also an effect along the \(x\) axis,  as emphasized by \cite[][see  their table 3]{Kruijssen+2019a}. In the rotational frame, the velocity dispersion at scale $R$ is thus:
\begin{equation}
    v_r=\nabla v_r  \, r\simeq 0.7\left(\frac{R}{1\text{pc}}\right) \,\kms,
    \label{eq-shear-vel}
\end{equation}
which is smaller by at least a factor of 3 than the turbulent velocity dispersion (see \S\ref{sec-CMZ}). The associated energy per unit mass can be estimated as:
\begin{equation}
    \frac{1}{2}v_r^2=\frac{1}{2}\left(\frac{3-\xi}{2}\right)^2\frac{GM(r)}{r^3}R^2.
\end{equation}
%where \(M(r)\propto r^\xi\) is the mass contained within the CMZ at radius $r$ as before and \(L_c\) is the size of the cloud. 
The collapsing barrier in HC08 thus becomes:
\begin{equation}
    e^{s_{c}}=\frac{2}{I}\frac{1}{\tilde{c}^2}\left[(1+\mathcal{M}_\star^2\tilde{a}^{2\eta})+\left(\frac{3-\xi}{2}\right)^2\frac{\mu}{3}\tilde{c}^2\right],
    \label{eq-shear}
\end{equation}
where $\mu=M(r_0)/\bar{\rho}r_0^3\simeq 0.08$. 

% {\bf *******************************}

% If the shear is not artificially enhanced, the barrier is raised by a factor about 2. From eq. \ref{correction_slope}, we can estimate a correction of 0.2 on the slope, using the scaling of Krieger, corresponding with \(\sigma\simeq3\) (the correction will be only 0.1 for the other scaling). With a Jeans mass of \(M_J=400 M_\odot\), we can estimate from \ref{slope} a slope of:
% \begin{equation}
%     x=-1.5-1+\frac{1.5}{2\times 3}\ln(2)+\frac{1.5^2}{2\times 3}\ln(400/10)\simeq - 0.95
% \end{equation}
% This is consistent with the observed slope around 10 \(M_\odot\) (see fig. \ref{CMF_EC_Krieger})

% {\bf *******************************}

\section{The cumulative role of shear and tides in the stability of an ellipsoidal deformable perturbation}
\label{app-tide-shear}

In the global energy budget of the perturbation, the shear is dominated by the tides because these latter act in 3 directions while the shear applies only on one axis. In order to quantify the global effect, we solve the complete system of the 3 virial equations accounting for both shear and tides. Starting from eqn.(\ref{Virial_eq_tides}) and adding a velocity dispersion coming from the shear along one of the axis (here the axis \(c\), the one for which the shear is the strongest relative to the tides), we have
\begin{align}
    \frac{1}{2}\frac{\partial^2 \tilde{c}^2}{\partial\tilde{t}^2} &=\frac{5}{2}(1+\mathcal{M}_\star^2 \tilde{a}^{2\eta})-\pi\alpha_1 \tilde{c}^2\e^s-\frac{\xi-2}{2}\tilde{c}^2\mu , \nonumber\\
    \frac{1}{2}\frac{\partial^2 \tilde{b}^2}{\partial\tilde{t}^2} &=\frac{5}{2}(1+\mathcal{M}_\star^2 \tilde{a}^{2\eta})-\pi\alpha_2 \tilde{c}^2\e^s-\frac{1}{2}\tilde{b}^2\mu+\frac{5}{2}\left(\frac{3-\xi}{2}\right)^2\mu \tilde{c}^2, \nonumber \\
    \frac{1}{2}\frac{\partial^2 \tilde{a}^2}{\partial\tilde{t}^2} &=\frac{5}{2}(1+\mathcal{M}_\star^2 \tilde{a}^{2\eta})-\pi\alpha_3 \tilde{c}^2\e^s-\frac{1}{2q_\phi^2}\tilde{a}^2\mu ,
\end{align}
where \(\mu=M_0/(r_0^3\bar{\rho})\) is the mass enclosed within a galactocentric radius $r_0=100$ pc, \(M(r)=r^\xi\) with \(\xi=2.2\), as inferred from the galactic density profile. By comparing the tidal terms and the shear term, we notice that the support   from both the shear and  the tides increases when \(\xi\to 0\) (\(\xi \ll 3\)), i.e. when the density quickly decreases. Note that the shear supports tangentially the cloud, explaining why the shear term enters in the equation of evolution of the axis \(b\).

\begin{figure}
    \centering
    \includegraphics[scale=0.7]{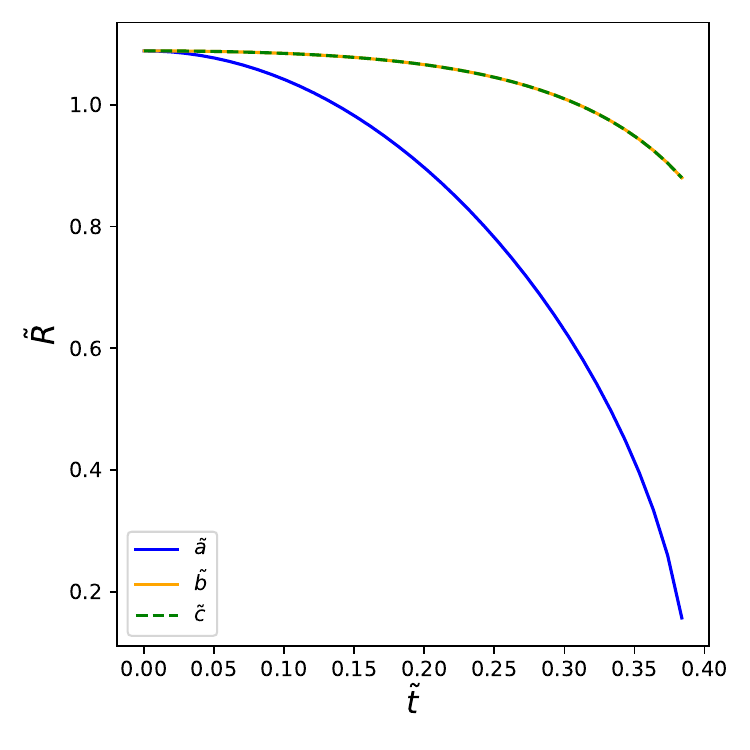}
    \caption{Evolution of the three axis of a 1 pc initially spherical density perturbation formed in a turbulence flow injected at 10 pc. The initial density is the HC08 critical density so in absence of tides and shear, the axis will not evolve. The turbulent velocity normalization is the one observed by Krieger et al. (2020). The time (x-axis) is normalized to the free fall time and the length (y-axis) to the Jeans length at the injection scale. The shear slows down the collapse but is unable to prevent it.}
    \label{Time_evol}
\end{figure}

In the case \(\xi = 2.2\), we have the equality:
\begin{equation}
    \frac{5}{2}\left(\frac{3-\xi}{2}\right)^2-\frac{1}{2} = -\frac{\xi-2}{2}<0.
\end{equation}
The evolution of the axis \(b\) and \(c\) will thus be the same  and the three axis will collapse if the initial density is the one given by the HC08 critical density in spite of the support of the shear. This confirms that the shear is dominated by the tides.
This is illustrated in Fig. \ref{Time_evol}.
% for a 10 pc perturbation.

% We solve this system of equations at the cloud scale. The evolution is illustrated in Fig. \ref{Time_evol} for a 10 pc perturbation. The initial density is the HC08 critical density. Then, during the evolution, the perturbation becomes less stable. The action of the shear and the tides is insufficient to stabilize the structure.
% \begin{figure}
%     \centering
%     \includegraphics[scale=0.7]{Time_evolution.PNG}
%     \caption{Time evolution of the axis of an ellipsoidal perturbation in the most favorable situation for the shear to prevent the collapse. The shear slows down the collapse but is unable to prevent it.}
%     \label{Time_evol}
% \end{figure}
 %The collapse barrier is thus lower than the HC08 usual barrier, as predicted by the tides and the global energy budget. 

%%%%%%%%%%%%%%%%%%%%%%%%%%%%%%%%%%%%%%%

\bibliography{references-cmz}{}
\bibliographystyle{aasjournal}

\end{document}